\documentclass[a4paper,10pt,fleqn]{article}

\usepackage[english]{babel}
\usepackage{inputenc}
\usepackage{amsmath,amsfonts,amssymb,marvosym}
\usepackage{hyperref}
\usepackage{breakurl}
\usepackage{graphicx}
\usepackage{subfigure}
\usepackage{times}
\usepackage{pifont}
\usepackage{enumerate}
\usepackage[]{lipsum}
\usepackage[mathlines]{lineno}
\usepackage{blindtext}
\usepackage{color}
\usepackage{blindtext}
\usepackage{wrapfig}
\usepackage[textwidth=17cm,textheight=23cm]{geometry}

\definecolor{newred} {RGB}{221,24,31}
\definecolor{newblue} {RGB}{0,96,173}
\definecolor{newgreen}{RGB}{0,128,0}

\usepackage{draftwatermark}
\SetWatermarkText{\Huge arXiv - 26. Jun 2018}

\title{Influence of patient alignment on image quality provided by a C-arm flat-panel detector computer tomography: a phantom study}

\author{Babak Alikhani\footnote{Corresponding author:alikhani.babak$@$mh-hannover.de}~$^{1,2}$~, Frank Wacker$^{1}$~, Thomas Werncke$^{1}$ \\{\small $^{1}$ Institute for Diagnostic and Interventional Radiology, Hannover Medical School, Hannover, Germany} \\ {\small $^{2}$ Center of Radiology / Nuclear Medicine, DIAKOVERE gGmbH, Hannover, Germany}}

\date{26. Jun 2018}

\begin{document}
\maketitle

\begin{abstract}\noindent
\centering
\begin{minipage}{0.7\textwidth}
\textbf{Purpose}: The aim of this phantom study was to evaluate the influence of patient alignment on the image quality by a C-arm flat-panel detector computer tomography (CACT). \\ \\
\textbf{Materials and Methods}: An ACR phantom (American College of Radiology CT accreditation phantom, Model 464, Gammex-RMI, Middleton, Wisconsin) placed in two opposite directions along the z-Axis was imaged using a CACT (Artis\,Zee\,Q; Siemens Healthcare, Forchheim, Germany). Image acquisition was performed by three different image acquisition protocols using fixed X-ray tube voltages of 81, 102 and 125\,kVp. The images were reconstructed with four different convolution kernels, i.e. normal, sharp, soft and very soft. Image quality was assessed in terms of high contrast image quality using the modulation transfer function (MTF) and low contrast image quality by assessing the signal-to-noise ratio (SNR) and contrast-to-noise ratios (CNR) as well as reliability of density measurements. Furthermore, the dose intensity profiles parallel and perpendicular to the patient support were measured free-in-air. \\ \\
\textbf{Results}: The intensity profile of the CACT measured by the detector system free-in-air showed that the anode heel effect is not in the longitudinal direction to the z-axis. The image noises measured in Setup A for the air and bone inserts were systematically higher compared to those measured in Setup B, in average about 3\% and 4\% for the air and bone inserts, respectively. An opposite behavior has been observed for the polyethylene, water-equivalent and acrylic inserts. The corresponding image noises were in average about 4\%, 6\% and 2\% lower measured in Setup A compared to those measured in Setup B. SNR for all inserts behaves inversely to the image noise. \\ \\
\textbf{Conclusion}: The patient alignment has a minor influence on the image quality of CACT. This effect is not based on the X-ray anode heel effect. It is caused mainly. This effect is caused mainly by the non-symmetrical rotation of the CACT. 
\end{minipage}
\end{abstract}

\twocolumn

\section{Introduction}
\label{sec:Introduction}

C-arm flat-panel detector computer tomography (CACT) allows acquisition and reconstruction of CT-like images in a flat-panel angiography system providing cross-sectional information during an interventional procedure. In addition to the additional diagnostic value, CACT images can be fused with real-time fluoroscopy, thus considerably expanding the diagnostic and therapeutic options of an angiographic system [1]. Thereby, the use of CACT has been increased in many radiologic and neuroradiologic institutions [2-6]. Although the low-contrast performance of CACT is lower in comparison to standard diagnostic multi-detector CT (MDCT), the measurements at the same radiation exposure show an improvement of the spatial resolution of CACT compared to MDCT\,[7-9], there is an increasing demand to high quality low-contrast CACT images. Furthermore, it has been shown that quantitative measurements at the same region are sufficient reliable\,[10]. Due to the conical beam geometry used in CACT as well as in cone-beam CT (CBCT) with wide aperture angles, it is to be expected that the homogeneity of the radiation field and thus the image quality of the CACT and CBCT is significantly influenced by the heel effect of the X-ray tube anode. Especially CACT with lower X-ray tube power and lower filtration and the absence of bowtie-filters is prone to X-ray-beam inhomogeneities. If the angular distribution of an X-ray tube spectrum or the heel effect is along to the patient support in a CACT, the image quality might be affected by the patient alignment, and could have an impact on the reliability of quantitative measurements. Previous investigations demonstrated the artifacts, for example, intensity inhomogeneities or errors induced in quantitative densitometry, caused by the heel effect and its influence on the image quality in CBCT\,[3,11-13]. \\
To the best of our knowledge, the impact of the patient alignment on the image quality provided by a CACT has never been studied. Therefore, the aim of this study was to investigate the impact of patient alignment on physical parameters related to the image quality. For this purpose, an ACR phantom was imaged using a CACT in two opposite directions along the patient support (z-axis) with different image acquisition protocols.


\section{Materials and Methods}
\label{sec:MaterialsMethods}


\subsection{Phantom}
\label{subsec:Phantom}

The ACR phantom used in this study consists of a water-equivalent material and contains four modules. Each module has a diameter of 20\,cm with a length of 4\,cm. A sketch and an AP projection of an ACR phantom are shown in Figure\,1 (a) and (b).  The modules are designed to determine image quality parameters, such as CT number, low- and high-contrast resolution. Since the influence of the patient alignment can reach its maximal change at the end of the phantom, the first and the last modules of the phantom only were utilized in this work. These modules are described as follows: \\
Using the first module, module 1, the spatial resolution or high-contrast can be determined. This module has eight aluminum bar resolution patterns, i.e. 4, 5, 6, 7, 8, 9, 10 and 12\,lp/cm, which are illustrated in Figure\,1 (c). The depth of the patterns is 3.8\,cm along the z-axis.The CT numbers of different materials can be determined using the module 4, which contains five cylinders of materials with different densities. The materials and their densities are as follows: air (0 g/cm$^3$), bone (1.95 g/cm$^3$), polyethylene (0.94 g/cm$^3$), water-equivalent cylinder (1 g/cm$^3$) and acrylic (1.18 g/cm$^3$). Each cylinder has a diameter of 25\,mm and a length of 4\,cm, except for the water-equivalent cylinder with a diameter of 50\,mm. Figure\,1 (d) shows a CT-image of the module\,4.
\begin{figure}[!ht]
\centering
\includegraphics[width=0.999\linewidth]{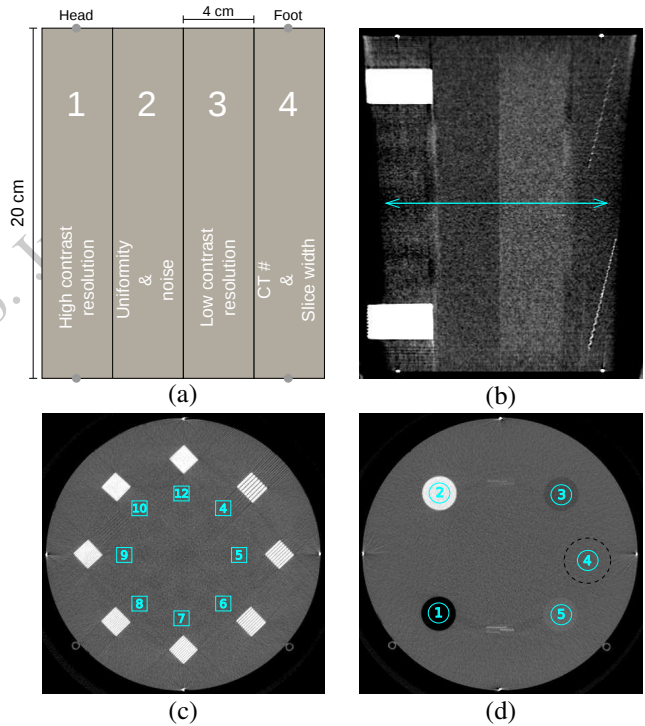}
\caption{(a) Sketch of an ACR phantom [14-16].  (b) ACR phantom imaged by the CACT. The solid (blue) arrow indicates the range of selected images for noise determination along the z-axis (see section Results).  (c) Module 1. The numbers in square present the count of line pairs per cm (lp/cm). (d) Module 4. The numbers indicate different materials: \ding{192} Air, \ding{193} Bone, \ding{194} Polyethylene, \ding{195} Water-equivalent material, \ding{196} Acrylic. The black, dashed ring indicates the water-equivalent cylinder.}
\end{figure}


\subsection{Image acquisition}
\label{subsec:Image_acquisition}

All measurements were repeated in two opposite positions of the phantom on the patient support. For Setup A the phantom was placed with the high contrast segment (module 1) feet first. CACT of the phantom was acquired using a monoplane, ceiling–mounted, angiographic system (Artis Zee Q; Siemens, Forchheim, Germany) using a CACT preset (6S DynaCTBody, Dyna CT; Siemens, Forchheim, Germany) with a 6s rotation run from ‑100° to 100° without additional filtering. Along this 200$^\circ$ rotation, 397 projections with a constant detector dose were acquired.\\
The CACT was equipped with a 30$\times$40\,cm flat-panel detector system [17,18]. The source-to-detector distance was 119.8\,cm. The measurements were performed using three different image acquisition protocols with X-ray tube voltages of 81, 102 and 125\,kV using the large focal spot (0.7 mm).\\
Table\,2 presents the image acquisition protocols (IAPs) of the measurements.
\begin{table} [!ht]
\centering
\small
\caption{IAPs of the measurements. \label{tab:IAPs}}
\begin{tabular}{cccc} \hline \hline
X-ray tube	&	Setup	&	X-ray tube	&	Dose Area Product	\\
voltage (kVp)	&		&	current (mA)	&	($\mu$Gy$\cdot$m$^2$)	\\ \hline
	81	&	A	&	253		&	1879.6			\\
	81	&	B	&	255		&	1892.0			\\
	102	&	A	&	91		&	1226.6			\\
	102	&	B	&	92		&	1232.7			\\
	125	&	A	&	39		&	1039.5			\\
	125	&	B	&	39		&	1045.1			\\ \hline \hline
\end{tabular}
\end{table} \\
The images were reconstructed with four different convolution kernels, normal, hard, soft and very soft. Afterwards the images were exported and analyzed using plug-ins and macros for the software program ImageJ (open-source image analysis software, version 1.50\,d; \burl{www.imagej.nih.gov/ij/}). 
The 2-D intensity profile of the X-ray tube was measured using 55\,kVp irradiating the detector free-in-air. 1-D profiles were calculated by averaging the 2-D profile along two axes parallel and perpendicular to the patient support. 1-D profile can be utilized to calculate the anode angle $\alpha$ of the X-ray tube. Thereby, the analytical model for heel effect described by Dixon \textit{et al.} [19,20] was used. The heel effect function $\rho(x)$ can be approximated as 
\begin{eqnarray} \label{eq:alpha}
\rho(x) \approx 1-\bar{\mu} \cdot d_0 \cdot \frac{x}{h \tan \alpha}\left( 1+\frac{x}{h  \tan \alpha}\right) ~,
\end{eqnarray}
where $\bar{\mu}$ is the weighted average of attenuation coefficients ($\mu$) over the X-ray spectrum, $h$ is the focal spot to fluorescent foil distance. \\
$\bar{\mu} \cdot d_0$ represents anode attenuation along the central ray of the X-ray beam. $\bar{\mu} \cdot d_0$ dependents on the X-ray energy and is empirically determined. For an anode angle of 7$^\circ$ and an X-ray tube voltage of 120 kVp is  $\bar{\mu} \cdot d_0 = 0.28$.

\subsection{Image Quality Assessment}
\label{subsec:Image_Quality_Assessment}

The high-contrast or spatial resolution of imaging systems was measured in terms of MTF by the use of the bar patterns in the module 1 of the phantom. Droege and Morin [21] devised a practical method for the MTF determination. This method was based on the standard deviation measurements of the pixel values within an image of bar patterns. The method introduced by Droege and Morin was used for the MTF assessment in this work. Therefore, 10$\times$10\,cm$^2$ square-shape region of interest (ROI) within bar patterns and the background were placed. For the MTF calculation an image set including 50 images were utilized. Using the CT numbers in the ROIs and their standard deviations, MTF for each spatial frequency (lp/cm) were calculated. \\
For low contrast resolution the signal-to-noise (SNR) and contrast-to-noise (CNR) ratio was assessed.  SNR is defined as the ratio between the mean CT number and its standard deviation, i.e. 
\begin{eqnarray}
\mathrm{SNR} ~=~\frac {\mid \overline{\mathrm{CT}}\#\mid}{\sigma} ~. \label{eq:SNR_p}
\end{eqnarray}
CNR is defined as 
\begin{eqnarray} \label{eq:CNR}
\mathrm{CNR} &=& \frac {\mid\overline{\mathrm{CT}}\#_{\mathrm{insert}}-\overline{\mathrm{CT}}\#_{\mathrm{bg}}\mid}{\sigma_{\mathrm{bg}}} ~,
\end{eqnarray}
where $\overline{\mathrm{CT}}\#_{\mathrm{insert}}$  is the mean CT number within a ROI. $\overline{\mathrm{CT}}\#_{\mathrm{bg}}$ and $\sigma_{\mathrm{bg}}$ denote the CT numbers of the background and the corresponding standard derivation in a ROI with the same area. \\
In order to calculate SNR and CNR, the inserts installed in the last module (module 4) of the ACR phantom were used. CT numbers using ROIs with an area of about 250 mm$^2$ were calculated. The corresponding standard derivations were calculated in ROIs with the same area placed at the middle of the module 4. Thereby, mean averages by the use of 25 images were calculated. The used images were reconstructed with normal kernel. \\
The image noise parallel to the patient support (z-axis) was determined using ROIs with an area of about 300 mm$^2$ at the center of images along the ACR phantom. Thereby, the noise profile by selection of 250 slices for different X-ray tube peak voltages was calculated in the central part of the phantom. To avoid the impact of cone-beam artefacts on the results, images at the border regions of the phantom were deselected (see Figure 1, top right).  Furthermore, by the use of the uniform module (module 2) the image noise perpendicular to the patient support (x-axis) was also calculated. Thereby, mean averages using 25 images were calculate.


\section{Results}
\label{sec:Results}

Figure 2 shows the intensity profiles of the CACT. On the top of this figure the 2-D profile is depicted, while the middle and bottom plots present the profiles along (z-axis) and perpendicular (x-axis) to the patient support, respectively. The intensity profile on the z‑axis shows an intensity variation about 4\%. The slope of the 1-D profile along the x‑axis caused by the heel effect delivers an anode angle of $\alpha = (8.84 \pm 0.02)$ degrees. 
\begin{figure}[!ht]
\centering
\includegraphics[width=0.99\linewidth]{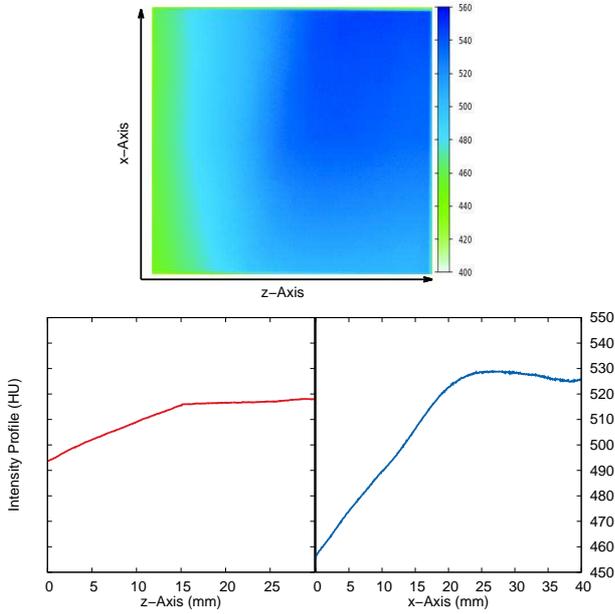}
\caption{2-D intensity profile of the detector (top). 1-D profile along the patient support (bottom, left) and perpendicular to the patient support (bottom, right). The profile along the x-axis indicates that the heel effect is perpendicular to the patient support. Using Equation (1) and 1-D profile along the x-axis, an anode angle of $\alpha = (8.84 \pm 0.02)$ degrees was determined. We presume that the kink in the profile along the z-axis is caused by the use of a grid between the X-ray tube anode and the patient support.}
\end{figure}
MTF values determined for both setups are shown in Figure 3.  Furthermore, using the Gaussian fit parameters the 50\%, 10\% and 2\% MTF for all reconstruction kernels were calculated, which are shown in Figure 4. \\
The related CT numbers of inserts with different densities for all X-ray tube voltages measured by setup A and B were calculated. As expected, the measured CT numbers behaved linearly relative to material density. Due to the beam-hardening effect of the photon fluence spectra, the CT numbers are decreasing with X-ray tube voltage and material density. Based on this effect, the largest difference of the CT numbers at the X‑ray tube voltages of 81, 102 and 125 kVp for bone insert was observed. \begin{figure}
\centering
\includegraphics[width=0.99\linewidth]{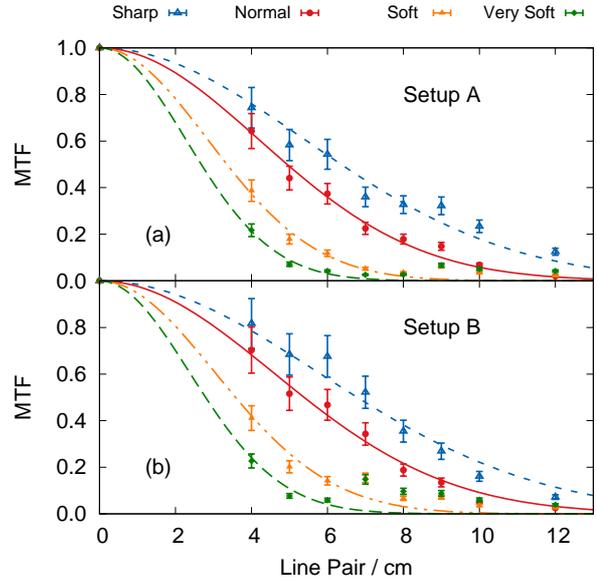}
\caption{MTF calculated for different convolution kernels for an X-ray tube voltage of 102 kV. (a) Measured for setup A (b) for setup B.}
\end{figure}
\begin{figure}
\centering
\includegraphics[width=0.99\linewidth]{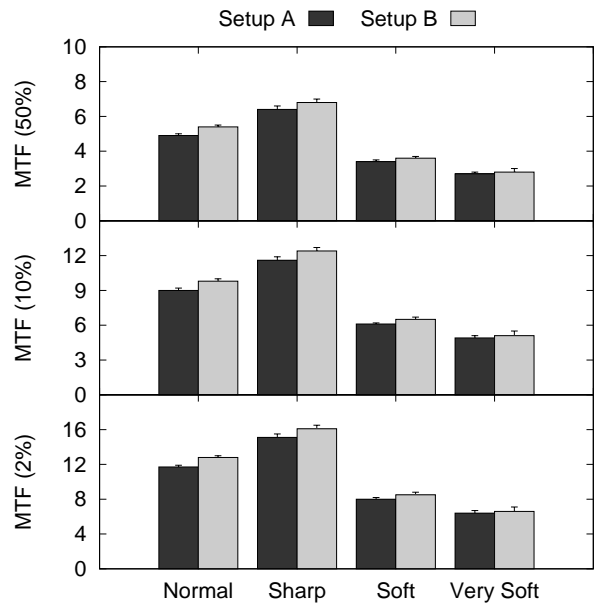}
\caption{50\%, 10\% and 2\% MTF for all reconstruction kernels for an X-ray tube voltage of 102 kV. The values are given in unit of line pair per cm (lp/cm).}
\end{figure}
For quantifying the effect of the phantom alignment on the CT numbers, the CT numbers of the inserts in both setups were determined and presented in Table 3. A considerable difference of the CT numbers measured by Setup A and Setup B was not be observed.
\begin{table*} [!ht]
\centering
\small
\caption{CT numbers of different materials measured by setup A (a), setup B (b).}
\begin{tabular}{lcccccc} \hline \hline
		&\multicolumn{2}{c}{80 kV}	&\multicolumn{2}{c}{102 kV}	&\multicolumn{2}{c}{125 kV}		\\ \hline
		&	Setup A	&	Setup B	&	Setup A	&Setup B	&	Setup A	&	Setup B		\\
Air		&$(-870 \pm 4)$	&$(-880 \pm 4)$	&$(-882 \pm 5)$	&$(-893 \pm 5)$	&$(-900 \pm 5)$	&	$(-914 \pm 5)$	\\
Polyethylene	&$(-110 \pm 4)$	&$(-109 \pm 4)$	&$(-134 \pm 5)$	&$(-127 \pm 5)$	&$(-192 \pm 5)$	&	$(-181 \pm 5)$	\\
Water-equivalent&$(10 \pm 5)$	&$(4 \pm 5)$	&$(-37 \pm 5)$	&$(-32 \pm 5)$	&$(-112 \pm 5)$	&	$(-99 \pm 6)$	\\
Acrylic		&$(88 \pm 6)$	&$(85 \pm 5)$	&$(57 \pm 6)$	&$(58 \pm 6)$	&$(-12 \pm 6)$	&	$(-5 \pm 6)$	\\
Bone		&$(1127\pm7)$	&$(1163\pm6)$	&$(930 \pm 7)$	&$(948 \pm 6)$	&$(724 \pm 7)$	&	$(743 \pm 6)$	\\  \hline \hline
\end{tabular}
\end{table*}
The image noises within the same ROIs for all material are presented in Figure 5.  The image noises measured in Setup A for the air and bone inserts were systematically higher compared to those measured in Setup B, in average about 3\% and 4\% for the air and bone inserts, respectively. An opposite behavior has been observed for the polyethylene, water-equivalent and acrylic inserts. The corresponding image noises were in average about 4\%, 6\% and 2\% lower measured in Setup A compared to those measured in Setup B. Figure 6 shows the SNR values.  As expected, SNR for all inserts behaves inversely to the image noise. This can be explained by use of Equation (2) and an almost constant CT-numbers of the inserts measured in setups A and B, which already presented in Table 3. In Figure 7 are the CNR values for five inserts shown. Except to the water-equivalent insert, the CNR values for all inserts calculated in Setup A are in average higher about 4\% than those calculated in Setup B.  CNR for water-equivalent insert is small with corresponding large uncertainties. It should be mentioned that the values in Figures 5,6 and 7 are normalized to the maximum of each histogram.
\begin{figure}[!ht]
\centering
\includegraphics[width=0.9\linewidth]{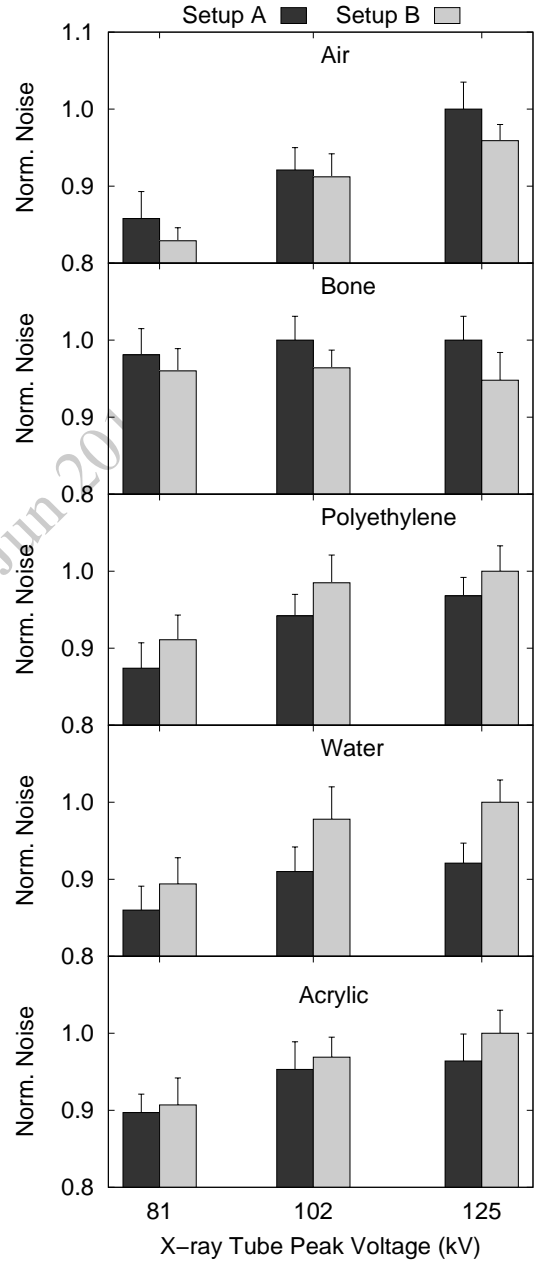}
\caption{ Normalized Noise calculated in five inserts. }
\end{figure}
\begin{figure}[!ht]
\centering
\includegraphics[width=0.9\linewidth]{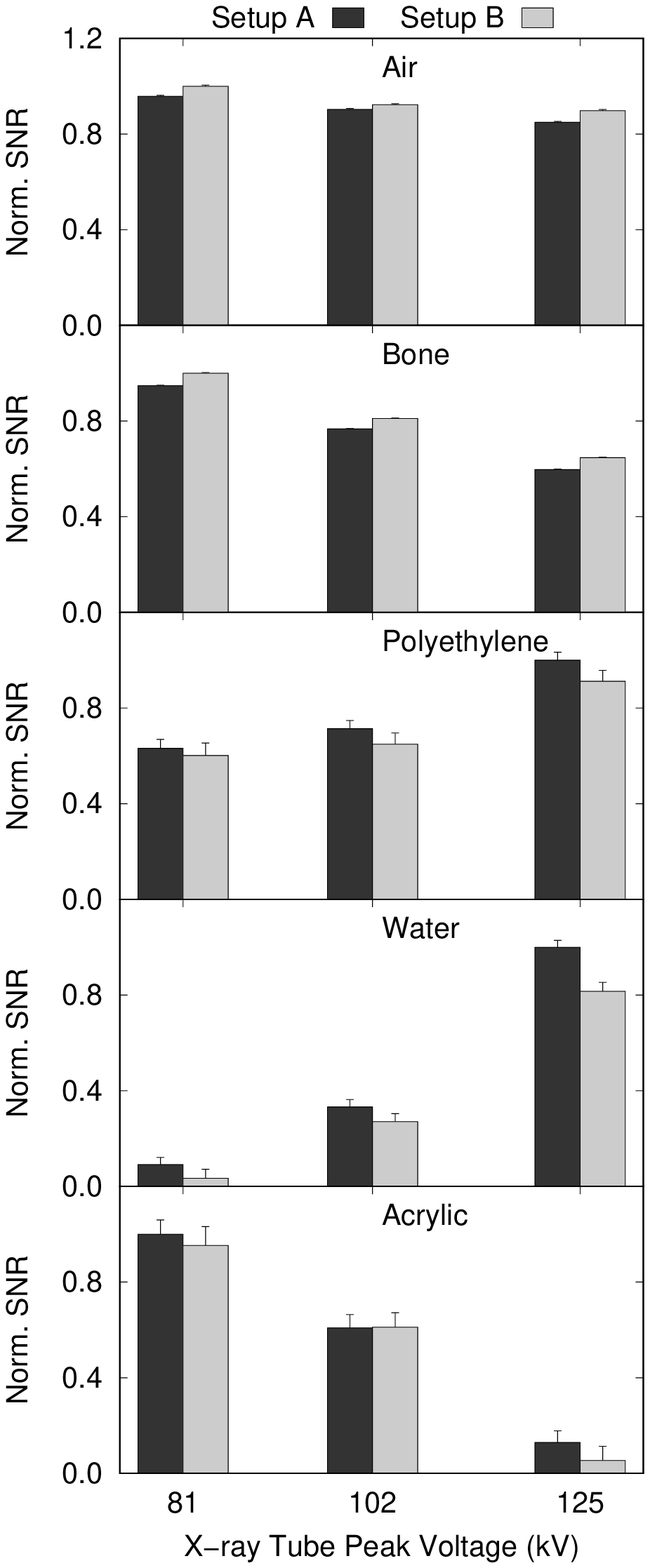}
\caption{ Normalized SNR calculated in five inserts. }
\end{figure}
\begin{figure}[!ht]
\centering
\includegraphics[width=0.9\linewidth]{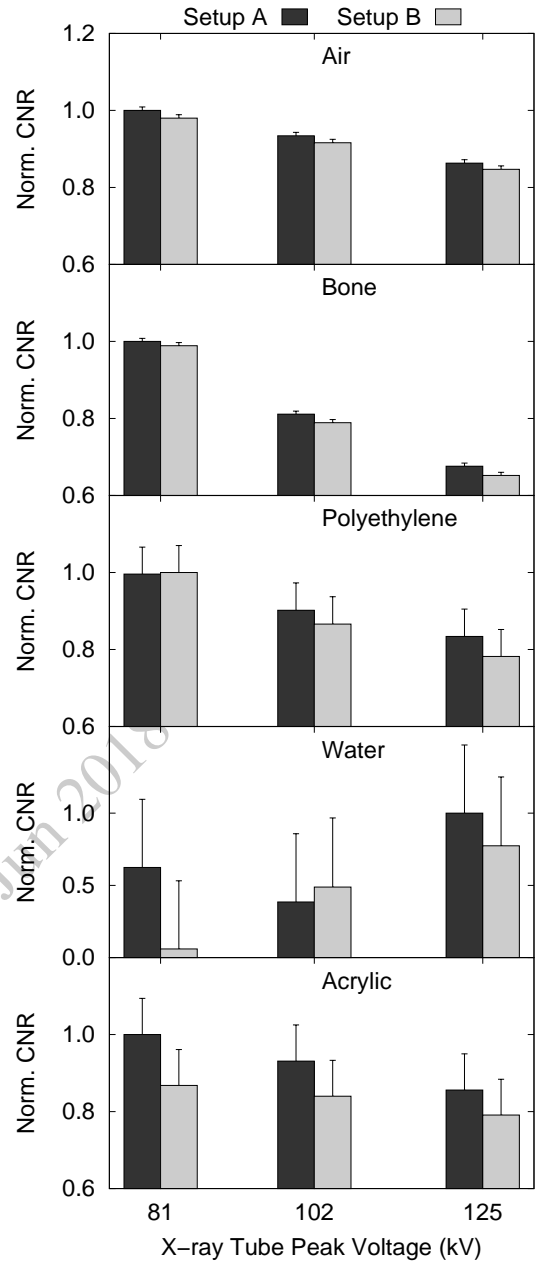}
\caption{ Normalized CNR calculated in five inserts. }
\end{figure}
The noise profiles parallel and perpendicular to the patient supports for both setups are presented in Figure 8. The staircase-shaped behavior of the noise plot measured along the z-axis is because of the different attenuation properties of the phantom modules. The radiation intensity through the uniform module has been less attenuated than through other modules, which is reflected in the gap of the noise diagram, Figure 8 (left). The image noise increasing at the middle of images along the x-axis is based on the cupping effect.
\begin{figure*}[!ht]
\centering
\includegraphics[width=0.9\linewidth]{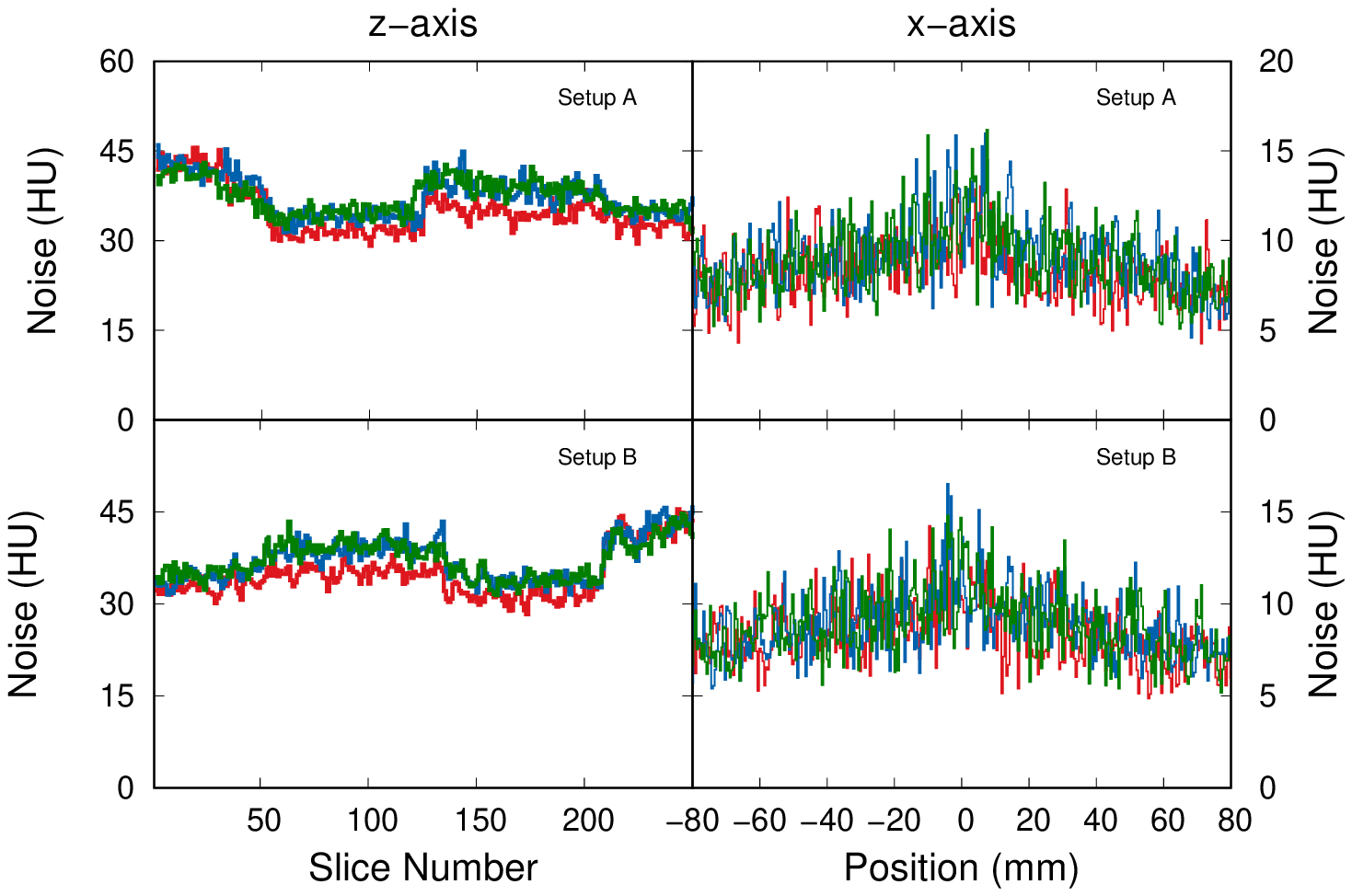}
\caption{Noise profile along the z-axis (left) and x-axis (right). The noise determination was performed by the setup A (top) and setup B (bottom). The red, blue and green lines indicate the noise values for the X-ray tube peak voltages of 81, 102 and 125 kV, respectively. The image noise increasing at the middle of images along the x-axis is based on the cupping effect.}
\end{figure*}


\section{Discussion}
\label{sec:Discussion}

The presented investigation demonstrated a negligible impact of patient alignment on the image quality of CACT. In addition, absolute density measurements with the CACT are still not feasible, so a conventional CT cannot be completely replaced by a CACT. \\
The 2D-profile measurement showed that the change of the image quality parameters, such as MTF, SNR and CNR, along the patient support (z-axis) was not based on the heel effect of the X-ray tube anode.  The intensity profile of the CACT measured by the detector system free-in-air showed an intensity gradient perpendicular to the patient support (x-axis). This finding indicates that the anode heel effect is not in the longitudinal direction to the z-axis. \\
Due to small varying of the radiation dose, given by dose area product, ranging from 0.5\% to 0.7\% for all X-ray tube voltage, the difference of the image quality parameters, such as image noise, SNR and CNR, measured by both setups cannot be attributed to the radiation dose. \\
Thereby, the observed differences of image quality, dependent on the phantom alignment are mainly caused by the incomplete rotation of the CACT around the z-axis. This incomplete rotation leads to an inhomogeneous irradiation of the object as compared to a conventional CT. This inhomogeneous dose distribution in the object/phantom results in an asymmetrical beam hardening and scattering [10,22-24] which leads to the orientation dependent image quality, observed in this study. This assertion was clearly noted by the measured image noises for the bone (+4\%) and acrylic (-2\%) inserts as the noise values showed a reversed variation by the setup A and B. The same but inverse effect could be observed by the air (+3\%) and polyethylene (‑6\%) inserts. \\
Furthermore, the image noise increase at the middle along the x-axis is attributed to the cupping effect [25-27]. We assert that the difference of the image quality parameters is also based on the local dose that reaches the end of the phantom by both directions. Since the dose measurement by the use of the ACR phantom was not possible, we focused our analysis on the noise determination along the z-axis, which reflects the local radiation dose. Both the intensity and noise profile along the z-axis validated the different dose along the patient support. The noise profile along the z-axis shows a reverse trend for both setups. \\
Up to now, in previous studies the low- and high-contrast performance of CACT comparing to MDCTs has been shown, however, there is no study, in our knowledge, demonstrating a relationship between patient alignment and the image quality in a CACT. \\
Bai \textit{et al.} [7] assessed the image quality of a CACT (Axiom Artis dTA; Siemens Healthcare, Forchheim, Germany) to that of two MDCTs (Lightspeed VCT; GE, Milwaukee, USA and Sensation Cardiac 64; Siemens Healthcare, Erlangen, Germany). Thereby, a male Anderson Radiation Therapy 200 phantom was used and the radiation dose with embedded thermoluminescence dosimeters (TLDs) was measured. The result indicated that the CACT applied fewer doses to the phantom by similar spatial resolution and low contrast detectability to both MSCTs. In other phantom and cadaveric study, Werncke \textit{et al.} [9] compared the effective radiation dose and image quality between a CACT (Artis Zee Q; Siemens, Forchheim, Germany) with a standard 16-slice MDCT (GE Lightspeed 16; GE Healthcare, Waukesha, Wisconsin). The radiation dose was determined using 100 TLDs placed in an anthropomorphic whole body phantom (adult male phantom with arms, model 701 and model 701-10, CIRSinc, Norfolk, USA). The result of this work also showed an improvement of the spatial resolution by the CACT in comparison to that of the MDCT at the same radiation dose. However, the low-contrast resolution, in terms of SNR, showed to be superior to MSCT. \\
Our study has a number of limitations: First, we used a cylinder-shaped phantom with a radius of 20 cm which is only an approximation for a patient body. Second, for obtaining the 2-D and subsequently 1-D intensity profiles and subsequently the calculation of the anode angel, we irradiated the detector system, which has an energy dependency. Using another X-ray tube voltage, the anode angle would be differing from 8.8 degrees. However, the main limitation is that the angular radiation dose distribution was missing. In order to obtain this information, a dose measurement using TLDs placed on the surface of the phantom had to be performed. This is time consuming and would in our opinion not alter the obtained results.  \\ \\
In conclusion, the patient alignment has a minor influence on the image quality in a CACT. This effect is caused mainly by the non-symmetrical rotation of the CACT.


\section*{References} \noindent \sloppy

[1] M Smyth, D Sutton, J Houston. Evaluation of the quality of CT-like images obtained using a commercial flat panel detector system. Biomed Imaging Interv J. 2, 2006.

[2] A Dörfler, T Struffert, T Engelhorn, \textit{et al.} Rotational flat-panel computed tomography in diagnostic and interventional neuroradiology. Fortschr Röntgenstr, 180, 2008.

[3] H Braun, Y Kyriakou, M Kachelrieß, \textit{et al.} The influence of the heel effect in cone-beam computed tomography: artifacts in standard and novel geometries and their correction. Phys Med Biol, 55, 2010.

[4] K A Hausegger, M Fürstner, M Hauser, \textit{et al.} Clinical Application of Flat-Panel CT in the Angio Suite. Fortschr Röntgenstr, 183, 2011.

[5] J H Siewerdsen. Cone-Beam CT with a Flat-Panel Detector: From Image Science to Image-Guided Surgery. Nucl Instrum Methods Phys Res A, 648, 2011.

[6] A Doerfler, P Gölitz, T Engelhorn, \textit{et al.} Flat-Panel Computed Tomography (DYNA-CT) in Neuroradiology. From High-Resolution Imaging of Implants to One-Stop-Shopping for Acute Stroke. Clin Neuroradiol, 25, 2015.

[7] M Bai, B Liu, H Mu, \textit{et al.} The comparison of radiation dose between C-arm flat-detector CT (DynaCT) and multi-slice CT (MSCT): A phantom study. Euro J Radiol, 81, 2012.

[8] S Ott, T Struffert, M Saake, \textit{et al.} Influence of different reconstruction parameters in the visualization of intracranial stents using c-arm flat panel CT angiography: experience in an animal model. Acta Radiol, 57, 2015.

[9] T Werncke, L Sonnow, BC Meyer, \textit{et al.} Ultra-high resolution C-Arm CT arthrography of the wrist: Radiationdose and image quality compared to conventional multidetector computed tomography. Euro J Radiol, 89, 2017.

[10] AK Jones, A Mahvash, Evaluation of the potential utility of flat panel CT for quantifying relative contrast enhancement, Med Phys, 39, 2012.

[11] S Mori, M Endo, K Nishizawa, \textit{et al.} Prototype heel effect compensation filter for cone-beam CT. Phys Med Biol, 50, 2005.

[12] B Li , TL Toth, J Hsieh, \textit{et al.} Simulation and analysis of image quality impacts from single source, ultra-wide coverage CT scanner. J Xray Sci Technol, 20, 2012.

[13] Y Kusano, S Uesaka, K Yajima, \textit{et al.} Positional dependence of the CT number with use of a cone-beam CT scanner for an electron density phantom in particle beam therapy. Radiol Phys Tech, 6, 2013.

[14] CT Accreditation Phantom Instructions, \burl{http://www.acraccreditation.org/~/ media/ ACRAccreditation/ Documents/CT/ CT-Accreditation-Testing-Instructions.pdf}.

[15] CT Accreditation Program: Image Quality and Dose Measurements, \burl{https://www.aapm.org/meetings/03AM/pdf/9785-27333.pdf}.

[16] CH McCollough, MR Bruesewitz, MF McNitt-Gray, \textit{et al.} The phantom portion of the American college of radiology (ACR) Computed Tomography (CT) accreditation program: Practical tips, artifact examples, and pitfalls to avoid. Med Phys, 31, 2004.

[17] JB Hinrichs, C von Falck, MM Hoeper, \textit{et al.} Pulmonary Artery Imaging in Patients with Chronic Thromboembolic Pulmonary Hypertension: Comparison of Cone-Beam CT and 64-Row Multidetector CT. J  Vasc Inter Radiol, 27, 2016.

[18] JB Hinrichs, T Murray, M Akin, \textit{et al.} Evaluation of a novel 2d perfusion angiography technique independent of pump injections for assessment of interventional treatment of peripheral vascular disease. Inter J Cardiovasc Imag, 33, 2017.

[19] RL Dixon, MT Munley, and E Bayram. An improved analytical model for CT dose simulation with a new look at the theory of CT dose. Med Phys, 32, 2005.

[20] RL Dixon and JM Boone. Analytical equations for CT dose profiles derived using a scatter kernel of Monte Carlo parentage with broad applicability to CT dosimetry problems. Med Phys, 38, 2011.

[21] R T Droege and R L Morin. A practical method to measure the MTF of CT scanners. Med Phys, 9, 1982.

[22] R Grimmer, M Kachelrieß, Empirical binary tomography calibration (EBTC) for the precorrection of beam hardening and scatter for flat panel CT. Med Phys, 38, 2011.

[23] H Dang, JW Stayman, A Sisniega, \textit{et al.} Cone-Beam CT of Traumatic Brain Injury Using Statistical Reconstruction with a Post-Artifact-Correction Noise Model. Proc SPIE Int Soc Opt En,. 2015.

[24] H Arakawa, MP Marks, HM Do, \textit{et al.} Experimental Study of Intracranial Hematoma Detection with Flat Panel Detector C-Arm CT. Am J Neurorad, 29, 2008.

[25] AK Hunter and WD McDavid. Characterization and correction of cupping effect artifacts in cone beam CT. Dentomaxillofacial Radiol, 41, 2012.

[26] AK Nagarajappa, N Dwivedi, R Tiwari. Artifacts: The downturn of CBCT image. J Inter Soc of Prev Comm Dentis, 5, 2015.

[27] JF Barrett and N Keat. Artifacts in CT: Recognition and Avoidance. RadioGraphics, 24, 2004.


\end{document}